\documentclass[multphys,vecphys]{svmult}

\usepackage{makeidx}         % allows index generation
\usepackage{graphicx}        % standard LaTeX graphics tool
\usepackage{multicol}        % used for the two-column index
\usepackage[bottom]{footmisc}% places footnotes at page bottom
\makeindex             % used for the subject index
\begin{document}

\title*{Stellar and Ionized Gas Kinematics of Peculiar Virgo Cluster Galaxies}
\author{Juan R. Cort\'es\inst{1}
, Jeffrey D. P. Kenney\inst{2}\and Eduardo Hardy\inst{3}}
\institute{Departamento de Astronom\'{\i}a, Universidad de Chile
\texttt{jcortes@das.uchile.cl}
\and Astronomy Department, Yale University \texttt{kenney@astro.yale.edu}
\and National Radio Astronomy Observatory \texttt{ehardy@nrao.edu}}
\maketitle

\begin{abstract}

We present the results of the stellar and ionized gas kinematics of 13 bright
peculiar Virgo cluster galaxies. The stellar velocity field are mostly consistent
with a rotation pattern, but some of them shows interesting features such as; S-shaped stellar isovelocity contours in NGC 4064, and
signatures of kinematical distinct components in NGC 4429, and NGC 4698.
This latter galaxy and NGC 4424 exhibit
extremely low $(V/\sigma)^{*}$ values suggesting that these
galaxies are the result of mergers. 
The ionized gas velocity fields are more disturbed than the stellar velocity fields, displaying non-circular motions. Most galaxies in the sample
reveals kinematical signatures that can be associated to gravitational interactions such as; mergers or tidal
interactions, being specially clear in the ``truncated/compact" galaxies. Moreover, most of the sample galaxies show evidence for both gravitational
interactions, and ICM-ISM stripping. Thus the evolution of a significant fraction
of cluster galaxies is likely strongly impacted by both effects.

\end{abstract}

\section{Introduction: Morphological evolution of Cluster galaxies}
%\label{sec:1}

It is well known that the environment affects the morphological types
 of galaxies in clusters. Observational facts such as the
``morphology-environment'' relation, ``Butcher-Oemler
effect'', as well as the results of the MORPHS collaboration
 (Dressler et al. 1997),
suggest that galaxies in
clusters evolve morphologically, with spirals becoming lenticular and redder
 as the result of environmental effects. 

Several mechanisms have been proposed for driving galaxy evolution,
including processes
that affect the gas and not the existing stellar stellar content
(e.g., ICM--ISM stripping; Gunn \& Gott 1972; Schulz \& Struck 2001; Vollmer
et al. 2001),
%Intracluster medium - interstellar medium (ICM-ISM) stripping
%(Gunn \& Gott 1972; Nulsen 1982; Schulz \& Struck 2001; Vollmer
%et al. 2001; van Gorkom 2004), gas accretion, and starvation 
%(Larson, Tinsley, \& Caldwell 1980), and those
and those interactions (Toomre \& Toomre 1972) that affect both the star and
the gas 
(e.g., mergers; Hernquist 1992)
%tidal interactions and collisions (e.g., Moore et al. 1996),
%and tidal interaction between galaxies and the cluster as a whole
However, it is still
largely unknown which processes do actually occur and which among them are dominant in
driving the morphological evolution of cluster
galaxies.

Detailed studies of the stellar and ionized gas kinematics could help
 to discriminate between interacting processes.
With this purpose in mind, we make a study of the stellar and ionized
gas kinematics of thirteen peculiar Virgo cluster galaxies using integral-field
spectroscopy techniques. Peculiar galaxies are natural targets
for studying galaxy evolution, since they are ``snapshots'' of the galaxy
evolution process. Moreover, the Virgo cluster is an ideal place for making this kind
of studies, since it has a moderately dense ICM, is dynamically young with
on-going sub-cluster mergers and infalling galaxies, and it has a
significant population  of galaxies characterized by truncated star formation
morphologies (Koopmann \& Kenney 2004).

\section{The galaxy sample}

The sample consists in thirteen peculiar Virgo cluster galaxies spanning a variety of optical
morphologies.
Morphological selection was made using the R, and H$\alpha$ atlas of Virgo cluster galaxies of
Koopmann et al. (2001), whereas the kinematical selection made use of the published
H$\alpha$ kinematics on 89 Virgo cluster spirals by Rubin et al. (1999).
While the sample selection is not uniform, it is
designed to include bright Virgo spirals
whose peculiarities are most poorly understood, and to
include representatives of the different H$\alpha$ types identified
by Koopmann \& Kenney (2004).

\section{Morphology of Peculiar Virgo Cluster galaxies}

\begin{figure}
\centering
\includegraphics[height=5cm]{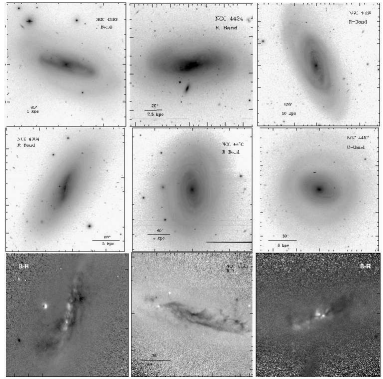}
%\picplace{5cm}{2cm} % Give the correct figure height and width in cm
%
\caption{Morphology of peculiar Virgo cluster galaxies. {\em Top row:}
galaxies with disturbed outer stellar isophotes ;NGC 4293, NGC 4424 and NGC 4569.
{\em Middle row:} barred and lensed galaxies; NGC 4064, NGC 4450, and NGC 4457.
{\em Bottom row:} disturbed dust distribution in NGC 4064, NGC 4293 and NGC 4424.}
\label{cortes:fig1}       % Give a unique label
\end{figure}

Optical imaging in the B, and R-band and H$\alpha$ narrow band was obtained
using the Kitt Peak WIYN 3.5m telescope with the Mini-mosaic imager.
A summary of the observed morphologies is displayed in figure $\ref{cortes:fig1}$. The results show that six galaxies of the sample display disturbed outer
stellar disks (NGC 4293, NGC 4351, NGC 4424, NGC 4569, NGC 4606 and NGC 4651), suggesting the action of gravitational
interactions. Signatures of triaxial structures such as bars and lenses are
found in NGC 4064, NGC 4450, and NGC 4457, whereas gross deviations from ellipticity
such as heart-shaped  isophotes, and non-elliptical isophotes are found in
NGC 4293, NGC 4424, NGC 4429, NGC 4606, and NGC 4694.

At least, six galaxies have disturbed dust distributions; NGC 4064, NGC 4293, NGC 4424, NGC 4569, NGC 4606, and NGC 4694. Most of them show disturbed outer
stellar disks, so they are probably the result of gravitational interactions.
On the other hand, three galaxies do not exhibit any optical evidence for
a recent gravitational interaction. These are NGC 4429, NGC 4457, and NGC 4580.

Most of the galaxies exhibit a depletion in their ionized gas, presenting
truncated H$\alpha$ distributions. NGC 4457, and NGC 4569 have truncated disks
and peculiar H$\alpha$ arms, suggesting an ongoing or recent ICM--ISM interaction.
NGC 4580 has a truncated H$\alpha$ disk with no peculiar H$\alpha$ arm or
H$\alpha$ asymmetry, but it does have strong stellar spiral arms in the
outer disk suggesting a  recent ICM--ISM interaction. In the rest of the
sample the cause of the depletion is not well understood with the present
data, although most of them could have suffered the action of ICM wind sometime
during their lifetime, as most of the galaxies in clusters, excepting the case
of NGC 4651, which displays a fairly normal H$\alpha$ disk.

\section{Stellar and Ionized gas kinematics}
We have mapped the stellar velocity fields, stellar velocity dispersion
fields, and ionized gas velocity fields using the DensePak Integral-field
unit installed at the 3.5m WIYN
telescope at Kitt Peak. Each galaxy was observed in the 4500 to 5000 {\AA},
allowing us to obtain the ionized gas kinematics using the H$\beta$ and
[O III]$\lambda\lambda$4959, 5007 lines, and the stellar kinematics using
the Magnesium triplet.

\begin{figure}
\centering
\includegraphics[height=8.0cm]{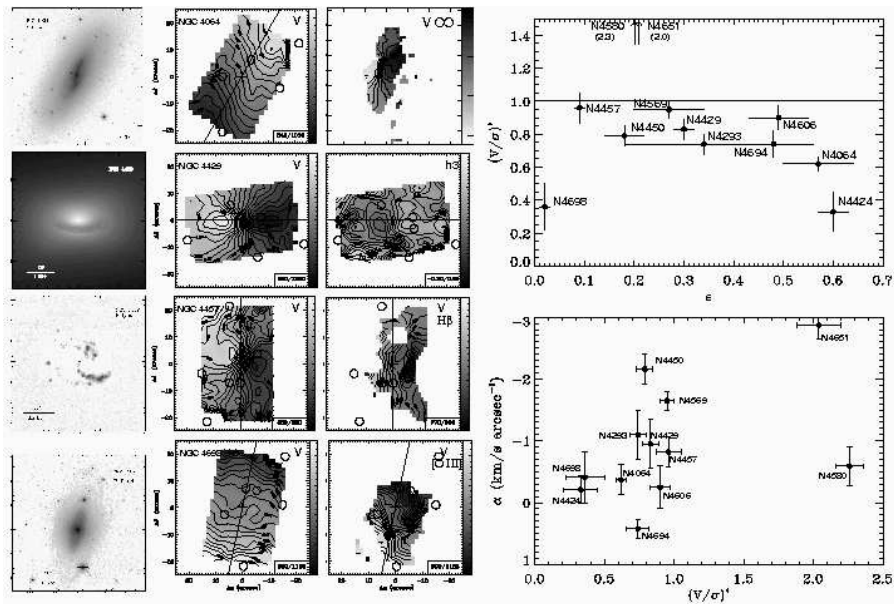}
%}
\caption{Stellar and Ionized gas kinematics. {\em Left
:} Stellar and ionized gas velocity fields for a set of sample galaxies.
{\em Right:} $(V/\sigma)^{*}$--$\epsilon$ diagram, and $(V/\sigma)^{*}$ versus
the slope of the velocity dispersion $\alpha$ for all the sample galaxies.}
\label{cortes:fig2}
\end{figure}

The stellar velocity fields in the galaxy sample are mostly consistent with
a rotational pattern, but they also exhibit a variety of interesting structures
(Figure $\ref{cortes:fig2}$).
Some galaxies display misalignments between the photometric and
kinematical major axes, which suggest the presence of non-axisymmetric structures,
starting from the clear kinematical misalignments in the stellar kinematics of NGC 4293
to the S-type shape found in the isovelocity contours of NGC 4064 as consequence of the existence of the stellar bar, which 
contrasts with the CO velocity field (Cort\'es et al. 2006; 
Figure $\ref{cortes:fig2}$) which displays strong non-circular
motions consistent with the infalling of gas to the center.

Signatures of a cold stellar disk (Figure $\ref{cortes:fig2}$) are found in NGC 4429, they characterize by  peaks
in velocity field correlated with peaks in $h_{3}$, and coincident with an undisturbed and axisymmetric dust disk, sugggesting that this cold stellar disk was probably formed by gas accretion.
Finally a remarkable twisting in the isovelocity contours is found in NGC 4698 (Figure $\ref{cortes:fig2}$) which corresponds
to a second kinematical component identified previously as an orthogonally rotating core by Bertola et al. (1999).

%The stellar velocity dispersion fields are as varied as the stellar velocity fields.
%Galaxies such as NGC 4450, NGC 4569, and NGC 4651 exhibit nearly symmetric velocity dispersion fields with a clear central peak.
%Other galaxies display more asymmetric stellar velocity dispersion fields such as; NGC 4064, NGC 4293,
%NGC 4424, and NGC 4429. Finally, galaxies such as NGC 4606, NGC 4694 and maybe NGC 4698 show
%constant velocity dispersion within the mapped region.

The ionized gas velocity fields, also exhibit a variety of structures, and in general look more disturbed than the stellar velocity fields. Non-circular motions are found in galaxies such as NGC 4064, NGC 4351, and NGC 4457.
In this last galaxy, their are specially important along the anomalous
H$\alpha$ arm (Figure $\ref{cortes:fig2}$), suggesting the action of ICM--ISM stripping
on a tilted gas disk (e.g; Schulz \& Struck 2001). 
Signatures of counter-rotation in the gas are found in NGC 4424, and evidence of a warped gas disk is found in NGC 4698 (Figure $\ref{cortes:fig2}$).

To understand the kinematics of the sample galaxies within the inner 2 kpc, we constructed a
$(V/\sigma)^{*}$--$\epsilon$ diagram  which is shown in
Fig. $\ref{cortes:fig2}$. 
This shows that NGC 4580, and NGC 4651 are systematically 
well above the oblate isotropic rotator line, showing
that their inner kinematics are dominated by the disk. 
On the other hand, NGC 4424, and NGC 4698 are extraordinary
cases, lying well below the line and indicating that these objects have more elliptical-like kinematics favoring a merger scenario. 
We also have found that  $(V/\sigma)^{*}$ correlates with the slope of the
slope of the stellar velocity dispersion $\alpha$ (Figure $\ref{cortes:fig2}$),
suggesting
that dynamically hot systems
tend to have flat velocity dispersion profiles in their inner 3 kpc.
The fact that some of these have small $C_{30}$, all the Truncated/Compact
galaxies defined by Koopmann \& Kenney (2004), and some with large $C_{30}$
suggest that the origin of this correlation should be due to an external cause,
in particular to mergers.

\section{Morphological evolution in Peculiar cluster galaxies}

Our results in the morphology and kinematics indicate that most of the
sample galaxies show evidence for both ICM-ISM stripping and gravitational
interactions. Two sample galaxies show evidence for recent ICM-ISM stripping,
but no strong recent gravitational interaction, just one galaxy show evidence
for a recent minor merger but no ICM-ISM stripping, and one does not show
any evidence of recent gravitational interaction or ICM-ISM stripping.

With the present evidence, results suggest that gravitational interactions
play an important role in altering the morphologies toward a more
lenticular stage (e.g. Bournaud et al. 2004) and driving gas inwards, but also drive gas
outwards in the form of tidal tails (e.g. Barnes \& Hernquist 1991), so
they are not responsible for the gas depletion in outer parts. ICM-ISM
stripping seems to play a key role in the depletion of gas in the
outer disk. Moreover, gravitational interaction can facilitate the action
of ICM wind over the ISM (e.g. NGC 4424). ICM-ISM
stripping plays an important role in ``pre-process'' spiral galaxies
in the core of the cluster, where merger or gravitational interactions
are more unlikely due to the high relative
velocities. Finally, these combined effects seems to be crucial
in objects with ``Truncated/Compact'' star formation as NGC 4064, and
NGC 4424.

%%%%%%%%%%%%%%%%%%%%%%%%%%%%%%%%%%%%%%%%%%%%%%%%%%%%%%%%%%%%%%%%%%%%%%  }

%%%%%%%%%%%%%%%%%%%%%%%%%%%%%%%%%%%%%%%%%%%%%%%%%%%%%%%%%%%%%%%%%%%%%%

\printindex

\begin{thebibliography}{99.}

\bibitem{journal} J. E. Barnes, \& L. Hernquist: ApJ, \textbf{370}, L65 (1991)
\bibitem{journal} F. Bournaud, F. Combes, \& C. J. Jog: A\&A, \textbf{418}, L27 (2004)
\bibitem{journal} F. Bertola, E. M. Corsini, J. C. Vega-Beltr\'an et al: ApJ, \textbf{519}, L127 (1999)
\bibitem{journal} J. R. Cort\'es, J. D. P. Kenney, \& E. Hardy: AJ, \textbf{131}, 747 (2006)
\bibitem{journal} A. Dressler, A. Jr. Oemler, W. J. Couch et al: ApJ, \textbf{490}, 577 (1997)
\bibitem{journal} J. E. Gunn, \& J. R. Gott: ApJ, \textbf{176}, 1 (1972)
\bibitem{journal} L. Hernquist: ApJ, \textbf{400}, 460 (1992)
\bibitem{journal} R. A. Koopmann, J. D. P. Kenney, \& J. Young: ApJS, \textbf{135}, 125 (2001)
\bibitem{journal} R. A. Koopmann, \& J. D. P. Kenney: ApJ, \textbf{613}, 866 (2004)
\bibitem{journal} V. C. Rubin, A. H. Waterman, \& J. D. P. Kenney: AJ, \textbf{118}, 236 (1999)
\bibitem{journal} S. Schulz, \& C. Struck: MNRAS, \textbf{328}, 185 (2001)
\bibitem{journal} A. Toomre, \& J. Toomre: ApJ, \textbf{178}, 623 (1972)
\bibitem{journal} B. Vollmer, V. Cayatte, C. Balkowski et al: ApJ, \textbf{561}, 708 (2001)



\end{thebibliography}
\end{document}